\documentclass{appolb}
\usepackage{graphicx}
\begin{document}
\title{Decays of light mesons
triggered by chiral chemical potential
\thanks{Talk given on the conference "Excited QCD" (Sintra, Portugal,7 - 13 May 2017). The funding for this work was provided by the Spanish MINECO under project
MDM-2014-0369 of ICCUB (Unidad de Excelencia `Maria de Maeztu'), grant FPA2016-76005-C2-1-P and grant 2014-SGR-104(Generalitat de Catalunya)(A.A. and D.E.),
RFBR project 16-02-00348 (A.A. and V.A.) and by  SPbSU travel grant 11.41.414.2017 (A.A.). }%
}
\author{A.~A.~Andrianov$^{1,2}$, V.~A.~Andrianov$^{1}$, D.~Espriu$^2$,  A.~E.~Putilova$^1$, A.~V.~Iakubovich$^1$
\address{$^1$ Faculty of Physics, Saint Petersburg State University, Universitetskaya nab. 7/9, Saint Petersburg 199034,
Russia}
\address{
$^2$ Departament de F\'isica Qu\`antica i Astrof\`isica and Institut de Ci\'encies del Cosmos (ICCUB), Universitat de
Barcelona, Mart\`i Franqu\`es 1, 08028 Barcelona, Spain
}}
\maketitle
\begin{abstract}
Light meson ( $\pi,\ \sigma,\ a_0$) properties in the environment with chiral imbalance are analyzed with the help of meson effective lagrangian
associated with QCD. New spatial parity violating decays of scalar
mesons arise as a result of mixing of $\pi$ and $a_0$ mesons under influence of chiral charge density. The pion electromagnetic
formfactor gets an unusual parity-odd contribution. Pion
effective masses depend on their velocities and may vanish in flight. The possible determination of chiral
chemical potential in heavy ion collisions based on above mentioned
phenomena is discussed.

\end{abstract}
\bigskip

\section{ Chiral Imbalance after nuclear collisions} The properties of nuclear (quark) matter in heavy ion collisions are  subjects of great interest which is supported by several running and planned heavy ion collider programs
 \cite{1,2}. New phases
of QCD  can be tested in current and future
accelerator experiments on RHIC, SPS and LHC \cite{3} and soon on FAIR and NICA. A fireball generated in these
collisions brings new data for experimental and theoretical
studies of various phases of strong interactions.

Study of hadron correlations in non-central heavy-ion
collisions at RHIC \cite{4} and LHC \cite{6}  provide \cite{7,8,9} the clue for
understanding of the so called
"chiral magnetic effect" (CME) in the reactions for peripheral ion
collisions \cite{10} .

On the contrary, a gradient density of isosinglet pseudoscalar condensate can be formed
as a result of  large, "long-lived" topological fluctuations
 of gluon fields
 in the fireball in central collisions (see
~\cite{aaep} for details). To describe various effects of hadron
matter in a fireball with Local Parity Breaking (LPB), we must introduce the axial/
chiral chemical potential ~\cite{aaep}.
There are some experimental indications of an abnormal
dilepton excess in the range of low invariant masses and rapidities
and moderate values of the transverse momenta
 (see the review in~\cite{tser}),
which can be partially thought of as a result of LPB
in the medium (the details can be found in ~\cite{tmf}). In
particular, in heavy-ion collisions at high energies, with raising
temperatures and baryon densities, a metastable state can appear
in the fireball with a nontrivial topological/axial charge
 ~$T_{5}$, which is
related to the gluon gauge field ~$G_{i}$.

Its jump ~$\Delta T_{5}$ can be
associated with the space-time integral of the gauge-invariant
Chern-Pontryagin density:
\begin{eqnarray}
\Delta T_5&=&T_5(t_f)-T_5(0)=\frac{1}{16\pi^2}\int^{t_f}_0
dt\int_{\mathrm{vol.}}d^3x\,\mathrm{Tr} (G^{\mu\nu}\widetilde
G_{\mu\nu})
\label{eq21}
\end{eqnarray}
where the integration is over the finite fireball volume.

It is known that the divergence of isosinglet axial quark current $J_{5,\mu}=\overline q
\gamma_\mu\gamma_5 q$ is locally constrained via the relation of
partial conservation of axial current (PCAC), affected by the gluon anomaly~\cite{aaep}.
It allows to find the connection of a nonzero topological
charge with a non-trivial quark axial charge ~$Q^q_{5}$,
\begin{eqnarray}
\label{eq23}
&&\frac{d}{dt}(Q_5^q-2 N_f T_5) \simeq 2i\int_{\mathrm{vol.}}d^3x\,
\widehat m_q \overline q\gamma_5q,
\\
&&Q_5^q=\int_{\mathrm{vol.}}d^3x\,q^\dagger\gamma_5 q = \langle
N_{L}-N_{R}\rangle,
\end{eqnarray}
where $\langle
N_{L}-N_{R}\rangle$ stands for the vacuum averaged difference between left and right chiral densities of baryon number ({\it chiral imbalance}).
Therefrom  it follows that in the chiral limit the generation of non-zero topological
charge in a finite fireball volume is accompanied by creation of chiral imbalance. Then if for the lifetime  of fireball and the
size of hadron fireball of order $L=5-10$~fm
the induced topological charge is non-zero, $\langle \Delta T_5
\rangle\ne 0$, then it may be associated with a topological
 potential $\mu_\theta$.  Equivalently the axial charge can be controlled by an axial chemical
potential~$\mu_5$ ~\cite{aaep} . Thereby we have
\begin{equation}
\langle \Delta T_5 \rangle \simeq \frac{1}{2N_f} \langle
Q_5^q\rangle\, \Longleftrightarrow\,\mu_5 \simeq
\frac{1}{2N_f}\mu_\theta, \label{eq24}
\end{equation}
Thus adding to the QCD lagrangian the term $\Delta{\mathcal
L}_{\mathrm{top}}=\mu_\theta\Delta T_5$ or $\Delta{\mathcal
L}_q=\mu_5 Q_5^q$, we get the possibility of accounting for non-trivial
topological fluctuations ("fluctons") in the nuclear (quark) fireball.

\section{Effective meson theories in the presence of chiral imbalance.}
For the detection of LPB in the hadron
fireball one considers the effective lagrangian  describing mass spectra and decays of scalar/pseudoscalar mesons in a fireball carrying a chiral imbalance.

In the environment with chiral chemical potential $\mu_5$ the scalar sector can be described in the meson Lagrangian
\cite{efflag},
\begin{equation}
D_\nu \Longrightarrow  \bar D_\nu - i \{{\bf I}_q\mu_5 \delta_{0\nu}, \star \}=\partial_\nu -i A_\nu [Q_{em}, \star]- 2i{\bf I}_q\mu_5 \delta_{0\nu},
\end{equation}
where $A_\mu, Q_{em}$ are the electromagnetic field and charge respectively.

The axial chemical potential
is introduced as a constant time component of an
isosinglet axial-vector field.
An effective Lagrangian includes the lightest scalar degrees of freedom $\sigma$ and $a_0(980)$, the latter being mixed with its pseudoscalar chiral partners $\pi$.

The effective lagrangian of the extended $\sigma$ model with two flavors contains the following set of operators (see  a similar lagrangian in \cite{giacosa,efflag}),
\begin{eqnarray}
&&\mathcal L = \frac{1}{4}\,Tr\,(D_{\mu}H\,(D^{\mu}H)^{\dagger})\label{kinterm}\\
&&
+\frac{b}{2}\,Tr\,[ \,m(H\,+\,H^{\dagger})]
+\frac{M^{2}}{2}\,Tr\,(HH^{\dagger})
\\
&&-\frac{\lambda_{1}}{2}\,Tr\,\left[(HH^{\dagger})^{2}\right]
-\frac{\lambda_{2}}{4}\,[\,Tr\,(HH^{\dagger})]^{2}
+\frac{c}{2}\,(\det H+\,\det H^{\dagger}) + \mathcal L_{WZW}(U).\nonumber
\label{lagr_sigma}
\end{eqnarray}
In the chiral parametrization they contain two types of fields in $H = \xi\,\Sigma\,\xi$ - the scalar isosinglet and isotriplet as well as the pseudoscalar isotriplet,
\[\Sigma = \left(\begin{array}{cc}v+\sigma+a^{0}_{0} & \sqrt{2}a^{+} \\
\sqrt{2}a^{-}_{0} & v+\sigma -a^{0}_{0} \\
\end{array}\right);\,  U = \xi\xi = \exp\left\{ \frac{i}{f_{\pi}}\left(\begin{array}{cc}\pi^{0}\! & \sqrt{2}\pi^{+} \!\\
\sqrt{2}\pi^{-} \!&  -\pi^{0} \!\\
\end{array}\right)\right\},
\]
where $v$ is a v.e.v. of  isoscalar field in the minimum of effective potential.
From spectral characteristics of scalar mesons in vacuum one fixes the lagrangian parameters,
\(\lambda_{1}=16.4850\), \(\lambda_{2}=-13.1313\), \(c=-4.46874\times10^4 {MeV}^2\), \(b  =1.61594\times10^5 {MeV}^2\).

The lagrangian is supplemented by the Wess-Zumino-Witten term describing parity-odd vertices \cite{WZW}. They contain light pseudoscalar fields and covariant derivatives and therefore are sensitive to chiral imbalance. The relevant WZW vertices for our purpose are,
\begin{equation}
\frac{ie\,\mu_{5}N_{c}}{6\pi^{2}\,v^{2}}\,\epsilon^{\;\:5\sigma\lambda\rho}_{4}\,
A_{\rho}\partial_{\sigma}\pi^{+}\,\partial_{\lambda}\pi^{-}
-\frac{e^{2}N_{c}}{24\,\pi^{2}v}\,\epsilon^{\,5\nu\sigma\lambda\rho}\,
\partial_{\sigma}A_{\lambda}\partial_{\nu}A_{\rho}\pi^{0} .\label{WZW}
\end{equation}
\section{Mass spectrum with chiral imbalance}
 For lightest isotriplet pseudoscalar $\pi$ and scalar $a_0$ states
the piece of the effective Lagrangian that is bilinear in fields looks as follows
\begin{equation}
\mathcal L=\frac{1}{2}\,(\partial_{\mu}a_{0}^j)^2
+\frac{1}{2}\,(\partial_{\mu}\pi^j)^2
-\frac{1}{2}m^{2}_{a,\mbox{\footnotesize\rm bare}}\left(a^{j}_{0}\right)^2-
\frac{1}{2}m^{2}_{\pi, \mbox{\footnotesize\rm bare}}\left(\pi^{j}\right)^2-\,4\mu_{5}\,\dot \pi^{j}a^{j}_{0}
\label{lag2}
\end{equation}
  Due to the last term in (\ref{lag2}) new eigenstates arise from the mixture of scalars and pseudoscalars,
  \begin{equation}
a_0= C_{a \tilde a} {\tilde a} + C_{a\tilde\pi} \tilde\pi,\quad \pi= C_{\pi \tilde a}\tilde{a} + C_{\pi\tilde\pi}\tilde\pi ,\label{mixed}
\end{equation}
For large 3-momentum $\tilde\pi$ becomes massless and then  tachyonic \cite{efflag}.
Such a behaviour does not represent a serious physical obstacle
 as it can be checked that the energies remain positive and no vacuum instabilities appear.
For the same 3-momentum  the new scalar states $\sigma, \tilde a_0 $ prove to increase in masses.
We present the results for the evolution of effective masses of $\tilde \pi$ and $\sigma, \tilde{a}_0$
 with respect to 3-momentum.

\begin{figure}[!htb]
\centering
\includegraphics[scale=.62]{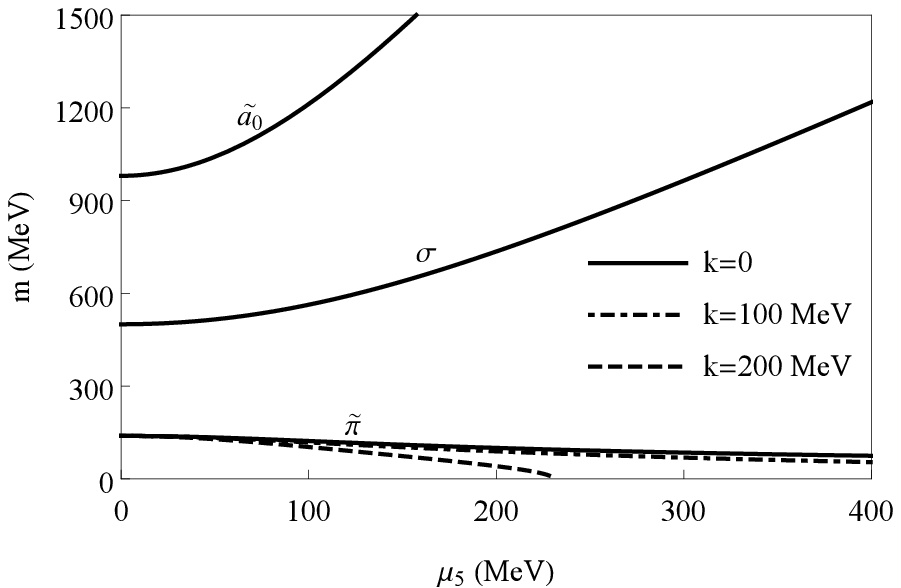}\quad
\label{fig:m1}
\includegraphics[scale=.45]{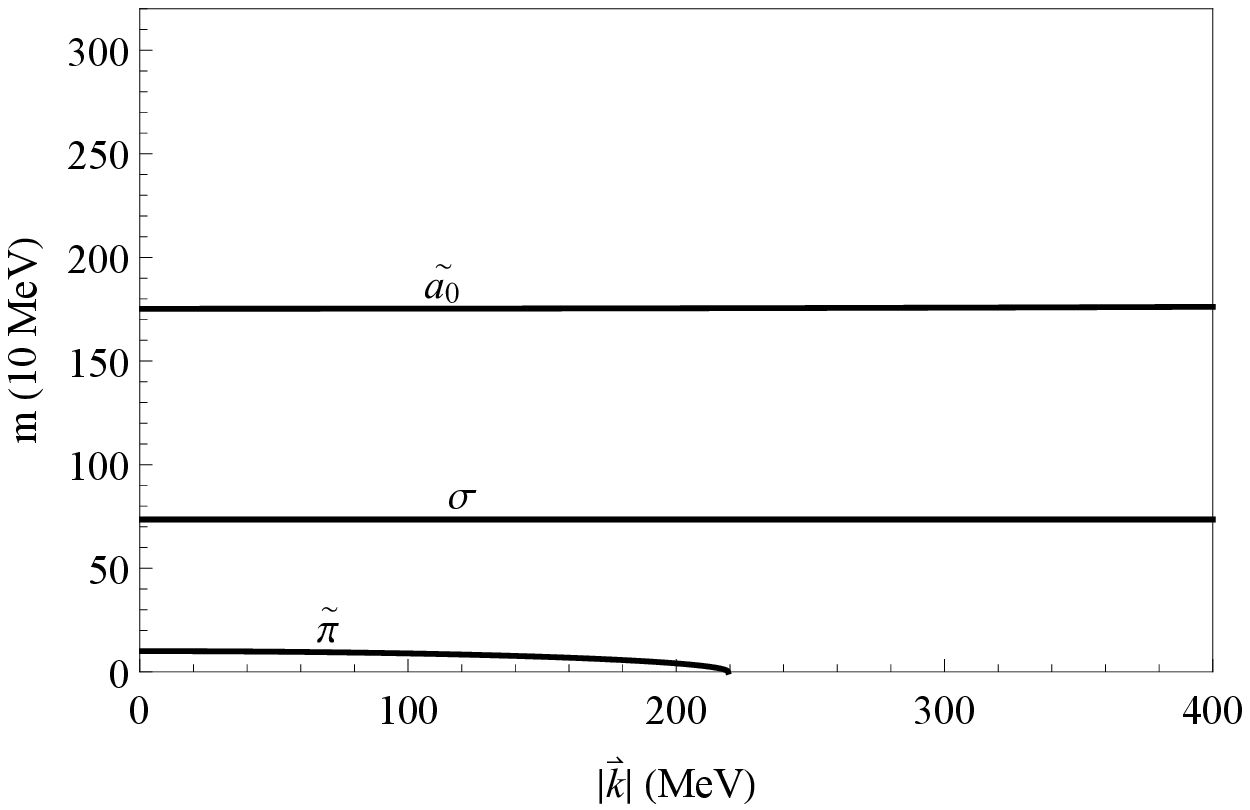}
\caption{Effective inflight masses at different values of 3-momentum and chiral chemical potential $\mu_5$}.
\label{fig:m^2_200}
\end{figure}

\section{ Exotic radiative decays in the enviroment with chiral imbalance}
The mixture of light pseudoscalars and heavy scalars (\ref{mixed}) in trilinear vertices, parity even (\ref{kinterm}) and parity-odd ones (\ref{WZW}), generate exotic decays of $\tilde a_0$ states \(a^{\pm}_0\rightarrow\pi^{\pm}\gamma\) which may serve as a "smoking gun" of strong parity breaking in a fireball produced in heavy ion collisions. The partial width of this decay is growing when chiral imbalance increases. As well the second vertex in (\ref{WZW}) induces the decay \(a^{0}_0\rightarrow\gamma\gamma\) which becomes stronger with increase of chiral imbalance.
 \begin{figure}[!htb]
\centering
\includegraphics[scale=.5]{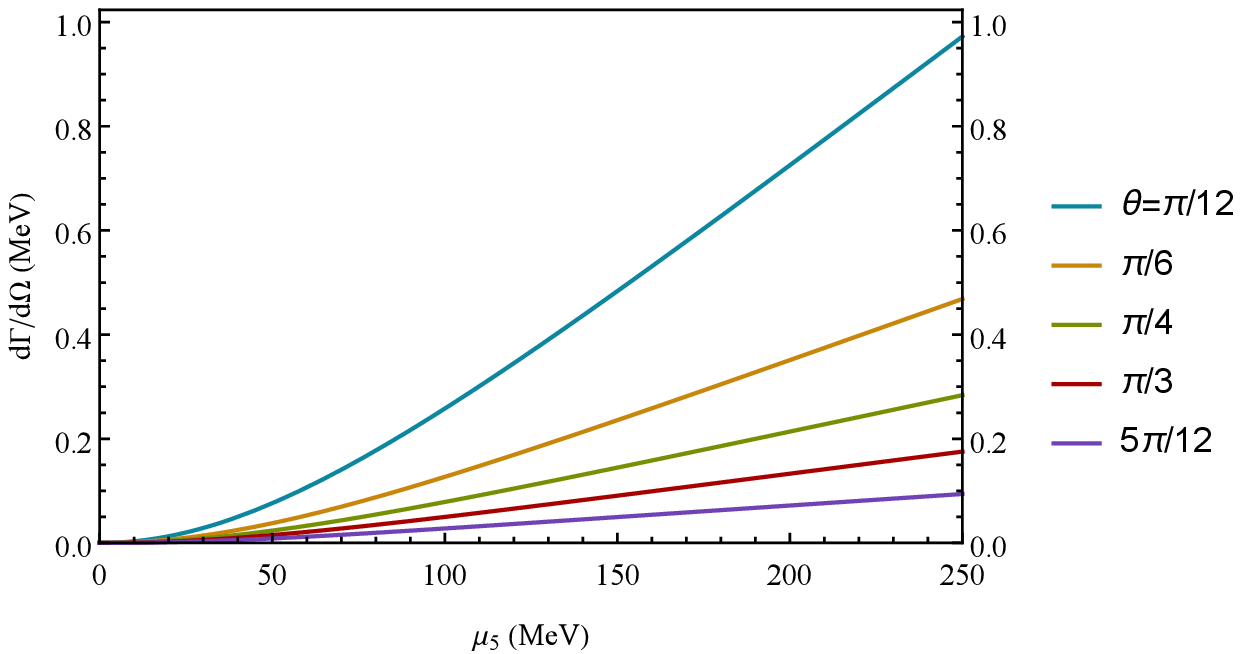}\quad
\includegraphics[scale=.45]{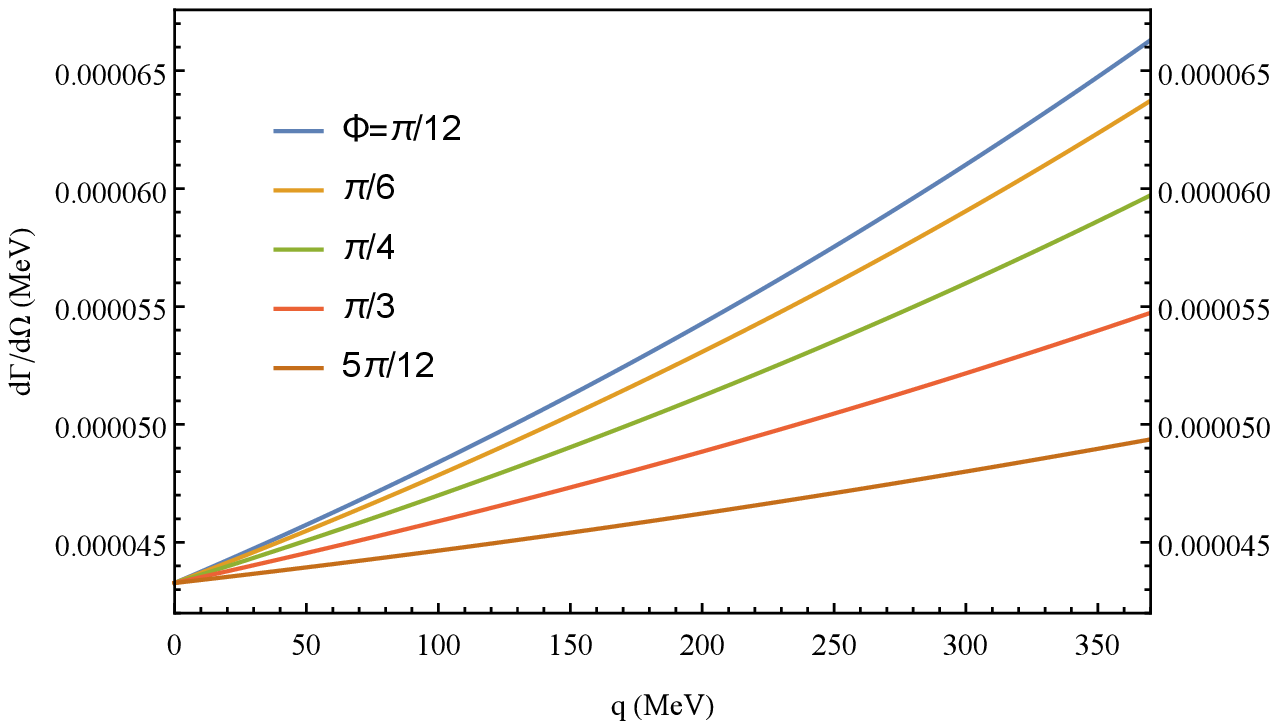}
\caption{
Decay widths of $\tilde a^\pm_0$ mesons at \(|\vec q|=200 MeV\) and of $\tilde a^0_0$ mesons at $\mu_5= 200 MeV$ in different angular sectors}
\label{fig4}
\end{figure}

Inverting the process \(a^{\pm}_0\rightarrow\pi^{\pm}\gamma\)  one may conclude that the exotic $a_0$ resonance decay in the intermediate state will considerably enhance the $\pi\gamma \to \pi\gamma$ scattering around 1 GeV in the medium with chiral imbalance.

\section{ Conclusions and outlook}
Strong CP violation is quite a challenging possibility
to be discovered in heavy-ion collisions both at high energy
densities (temperatures) and when triggered by large
baryon densities. For that purpose we suggest to detect
peculiar effects generated in CP odd background (chiral chemical potential),
measuring the probability of production of the scalar states
with  indefinite CP parity.

The influence of chiral imbalance on thermodynamics of hadron(quark) matter can be efficiently established in lattice computations by the chiral chemical potential method \cite{braguta}.

On the previous conference "eQCD2016"\cite{eqcd16} we declared also a manifestation for LPB in the presence of chiral imbalance in the sector of  $\rho$ and $\omega$ vector mesons.  It turns out\cite{aaep} that the spectrum of massive vector mesons splits into three components with different polarizations having
different effective masses $m^2_{V,+} < m^2_{V,L}< m^2_{V,-}$. Then a resonance broadening occurs that leads to an
increase of the spectral contribution to the dilepton production
as compared with the vacuum vector meson
states.  The latter
mechanism for generating local spatial parity breaking helps to (partially) saturate the anomalous yield of
dilepton pairs in the CERES, PHENIX, STAR, NA60, and ALICE
experiments.

We draw also attention to the recent interesting proposal to measure the photon polarization asymmetry in $\pi \gamma$ scattering \cite{harada} as a way to detect LPB due to chiral imbalance.


\end{document}